\documentclass[useAMS,usenatbib,usegraphicx]{mn2e}
\usepackage{amsmath}

\title[A high-sensitivity polarimeter]{A
high-sensitivity polarimeter using a ferro-electric liquid crystal modulator}
\author[J. Bailey et al.]{Jeremy Bailey$^{1,2}$\thanks{E-mail:
j.bailey@unsw.edu.au}, Lucyna Kedziora-Chudczer$^{1,2}$, Daniel V. Cotton$^{1,2}$, 
\newauthor Kimberly
Bott$^{1,2}$, J. H. Hough$^3$,
P. W. Lucas$^3$ \\
$^{1}$School of Physics, UNSW Australia, NSW 2052, Australia\\
$^{2}$Australian Centre for Astrobiology, UNSW Australia, NSW 2052, Australia.\\
$^{3}$Centre for Astrophysics Research, Science and Technology Research Institute, University of
Hertfordshire, Hatfield AL10 9AB, UK}
\begin{document}

\date{Accepted 5 March 2015 Received 8 February 2015; in original form 27 November 2014}

\pagerange{\pageref{firstpage}--\pageref{lastpage}} \pubyear{2015}

\maketitle

\label{firstpage}

\begin{abstract}
We describe the HIgh Precision Polarimetric Instrument (HIPPI), a polarimeter built at UNSW Australia
and used on the Anglo-Australian Telescope (AAT). HIPPI is an aperture polarimeter using a ferro-electric
liquid crystal modulator. HIPPI measures the linear polarization of starlight with a
sensitivity in fractional polarization of $\sim$4 $\times$ 10$^{-6}$ on low polarization objects and
a precision of better than 0.01\% on highly polarized stars. The detectors have a high dynamic range allowing
observations of the brightest stars in the sky as well as much fainter objects. The telescope
polarization of the AAT is found to be 48 $\pm$ 5 $\times$ 10$^{-6}$ in the g$'$ band.
\end{abstract}

\begin{keywords}
polarization -- instrumentation: polarimeters -- techniques: polarimetric.
\end{keywords}

\section{Introduction}

Stellar polarization can be measured to very high sensitivity with ground-based telescopes. Unlike
photometry where atmospheric effects limit the achievable precision, polarimetry is a
differential measurement and so there is no fundamental limit on the sensitivity. \citet{kemp87}
measured the polarization of the Sun to levels of parts in ten million, and astronomical polarimeters
have been built that measure stellar polarization at the parts per million level \citep{hough06,wiktorowicz08}. 
Interest in such instruments has been driven, in
particular, by the possibility of detecting polarized scattered light from hot Jupiter type exoplanets,
which is predicted to be at levels of $\sim10^{-5}$ or less in the combined light of the star and the
planet \citep{seager00,lucas09}. Such instruments also have other applications such as the study of
the local interstellar medium \citep{bailey10,frisch12} and the scattered light from debris disks
\citep{wiktorowicz10}.

The standard method used for most high sensitivity polarization studies has been the use of the
photoelastic modulator (PEM) technology pioneered by James Kemp \citep{kemp69,kemp81,kemp87}.
These devices modulate polarization at frequencies of 20 -- 100 kHz by means of the stress
birefringence in an optical material made to vibrate at its natural frequency using piezoelectric
transducers. While PEMs have proved very successful in this role they nevertheless have a number of
disadvantages. While a high modulation frequency is desirable to minimize the effects of variations
due to seeing and tracking errors, the frequency of PEMs is higher than is really needed for this 
purpose. The high frequency can present problems in providing a suitable detector system. Many
detector types can either not operate at the required speed, or can only do so with some compromise
in their noise performance. PEMs are inherently sine-wave modulators and thus the efficiency of a PEM
polarimeter is less than that of an ideal square wave modulator by a factor of at least $\sqrt{2}$
(actually slightly more than this, see \citealt{hough06}). PEMs are also quite bulky and thus difficult
to use where space is limited. 

In this paper we describe a high-sensitivity polarimeter using an alternate type of polarization
modulator, a ferro-electric liquid crystal modulator (FLC). FLCs are electrically switchable wave
plates consisting of a thin layer of liquid crystal material sandwiched between two glass plates.
They have a fixed retardation and the orientation of the optical axis can be controlled by an applied
drive voltage. FLCs have been used, in particular, for solar polarimetry 
\citep[e.g.][]{gandorfer99,martinez99,hanaoka04} and also for the Extreme Polarimeter
\citep[ExPo][]{rodenhuis12}, a high contrast imaging polarimeter and the SPHERE/ZIMPOL instrument for
the ESO VLT \citep{bazzon12,roelfsema14}.  

Our polarimeter HIPPI (HIgh Precision Polarimetric Instrument) has been used on the 3.9m
Anglo-Australian Telescope (AAT) at Siding Spring Observatory for two observing runs during 2014.
The design of HIPPI has been based on that of the previous PlanetPol instrument \citep{hough06}.
However, HIPPI differs from PlanetPol in being optimised for observation at blue wavelengths. This is
based on results such as those of \citet{pont13} and \citet{evans13} that indicate the presence of
strong Rayleigh scattering at blue wavelengths from the exoplanet HD 189733b making these wavelengths
the most suitable for detecting exoplanet polarization.

\section{Instrument Description}

\subsection{Overview}

\begin{figure}
\begin{center}
\includegraphics[scale=0.5, angle=0]{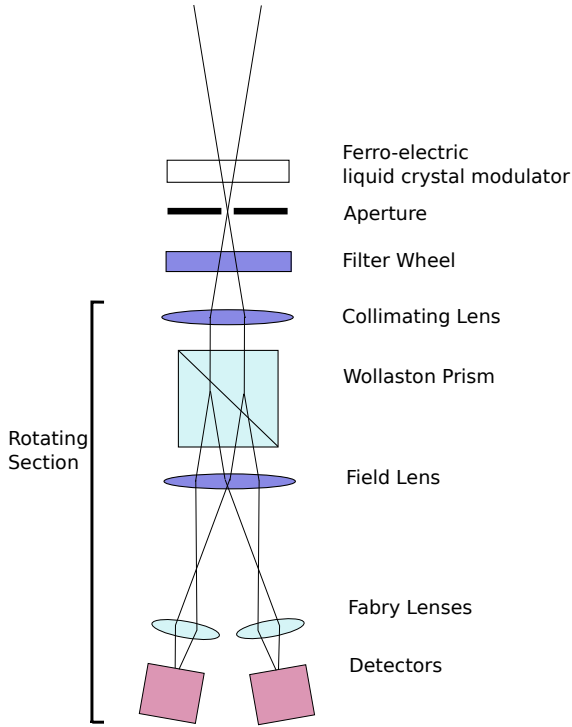}
\caption{Schematic diagram of HIPPI optical system (not to scale)}
\label{fig1}
\end{center}
\end{figure}

A schematic diagram of the HIPPI optical system is given in figure \ref{fig1}. The FLC modulator is
the first element in the optical system. This is an important design feature since any optics placed
ahead of the modulator could potentially induce spurious polarization effects, for example,
polarization due to inclined mirrors or residual stress birefringence in refracting elements.
Following the FLC is an aperture of 1 mm diameter corresponding to 6.7 arc seconds at the
AAT f/8 focus. This is followed by a six position filter wheel.

The filters used with HIPPI have been Sloan Digital Sky Survey \citep[SDSS,][]{fukugita96} g$'$ and r$'$ filters (from Omega Optical, giving wavelength ranges of $\sim$400
-- 550 nm and $\sim$550 -- 700 nm respectively). There is also a short pass filter which passes
wavelengths shorter than 500 nm (referred to as 500SP). The instrument has little throughput below about 350 nm due
to absoprtion in the calcite prism so the range of this filter is from $\sim$350 -- 500 nm. The filter wheel also
includes a clear position and a blank setting that can be used for
taking dark measurements.

The polarization analyser is a calcite Wollaston prism that provides a 20 degree beam separation. 
This is placed between two lenses to collimate the light through the prism. A Fabry lens in each beam
images the telescope pupil onto the two detectors. The whole optical system from the collimating lens
to the detectors is rotatable about the optical axis using a Thorlabs NR360S NanoRotator stage.
Rotating this system through 90 degrees relative to the modulator has the effect of reversing the
sign of the modulation seen by the detectors and provides a ``second stage chopping'' which helps to
improve accuracy by eliminating some systematic effects \citep{kemp81}. A similar system was used in
PlanetPol \citep{hough06}. All the optics are anti-reflection coated for the wavelength range 350 -- 700 nm. 

The instrument components are mounted on a standard 300 mm square aluminium optical breadboard that
is attached by 90 degree brackets to a mounting plate that bolts to the back of the telescope. Many
of the structural components, optical mounts and electronics enclosures have been constructed by 3D
printing in ABS plastic. The instrument is therefore compact and lightweight (10 kg).

\subsection{Ferro-electric liquid crystal modulators}

Two different FLC modulators have been used with HIPPI. The first is a LV1300-AR-OEM device from
Micron Technology\footnote{This company no longer supplies such devices}. It is designed for the
400--700 nm range and is 12.7 mm in diameter housed in a 25 mm diameter cell. The second is an MS
Series polarization rotator from Boulder Nonlinear Systems (BNS) designed for the wavelength range 425 --
675 nm and is 22 mm in diameter with a 15 mm useful aperture. Both devices are designed to be half-wave
retarders at a wavelength near 500 nm, and depart from half-wave away from this wavelength as discussed further in section
\ref{sec_bandpass}. 

The two modulators are very similar in their operation and provide good polarization modulation with a $\pm$5 V
drive waveform. However, we have found the BNS modulator to produce much lower levels of
instrumental polarization, and it is therefore currently the preferred option.

Electrically the modulators are equivalent to capacitors of $\sim$200 nF and therefore require a drive
circuit
capable of driving at high speed into a capacitive load. The devices can also be damaged by sustained DC
voltages. We have desiged and built a drive circuit consisting of a two-pole Butterworth high pass filter
followed by an amplifier using a NPN/PNP transistor pair output stage. The filter ensures no DC or low
frequency components reach the device. The drive amplifier has the high slew rate, and high drive
current needed to drive a square wave into the capacitive load.

The drive waveforms are generated in software.  A simple square wave between +5 V and $-$5 V 
has been used for all the observations described in this paper. Our system allows selection of 
modulation frequencies between 200 Hz and 2 kHz. We have found 500 Hz
to be a good choice for actual observing, providing a close to square wave modulation, while being fast
enough to be insensitive to intensity fluctuations due to seeing or tracking errors.

FLCs are temperature sensitive devices. The switching is faster at higher temperatures and the
switching angle is also temperature dependent. To ensure consistent and stable operation we mount the FLC in a
temperature controlled lens tube and operate it at a constant temperaure of 25$^\circ$ C, maintained to
about $\pm$0.1$^\circ$ C. 

\subsection{Detectors}

The detectors used in HIPPI are compact photomultiplier tube (PMT) modules. The modules contain 
a metal packaged photomultiplier tube combined with an integrated HT supply. PMTs have substantial advantages of
large detector area and low dark noise compared with possible solid-state alternatives such as avalanche
photodiodes (as used in PlanetPol) or so-called ``silicon photomultipliers''. 

HIPPI uses Hamamatsu H10720-210 PMT modules which have ultra-bialkali photocathodes 
\citep{nakamura10} providing a quantum efficiency (QE) of 43\% at 400 nm. They are compact modules operating off a single
5V supply. For the high photon rates required with HIPPI it is not possible to use the PMT in a photon
counting mode. Instead we use a transimpedance amplifier to amplify the photo-current. The amplifiers
designed and built for HIPPI use an ultra low noise Texas Instruments OPA 129 operational amplifier with an
extremely low input current noise of 0.1 fA Hz$^{-1/2}$. Remotely switchable transimpedance gains of
10$^5$, 10$^6$ and 10$^7$ V/A can be selected. The PMT module itself provides selectable HT voltages
from 500 V to 1100 V corresponding to a variation of the PMT gain from $5 \times 10^3$ to $3 \times
10^6$ electrons per photon. The ability to remotely vary both the photomultiplier gain and amplifier gain over a wide
range provides a very high dynamic range, allowing HIPPI to observe objects from the brightest stars
in the sky to quite faint objects while still providing close to photon noise limited performance. 

\begin{figure}
\begin{center}
\includegraphics[scale=0.4, angle=270]{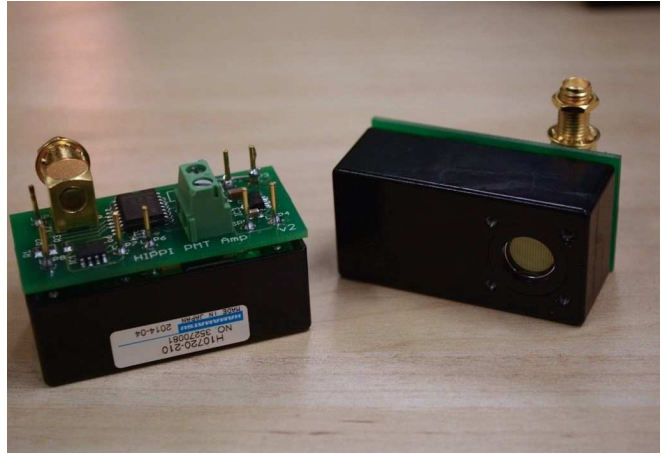}
\caption{HIPPI Detector Modules}
\label{fig2}
\end{center}
\end{figure}

The PMT amplifiers have been built in surface-mount construction on a compact printed circuit board
25 $\times$ 50 mm that fits on the back of the PMT module as shown in figure \ref{fig2}.  

\begin{figure*}
\begin{center}
\includegraphics[scale=0.8, angle=0]{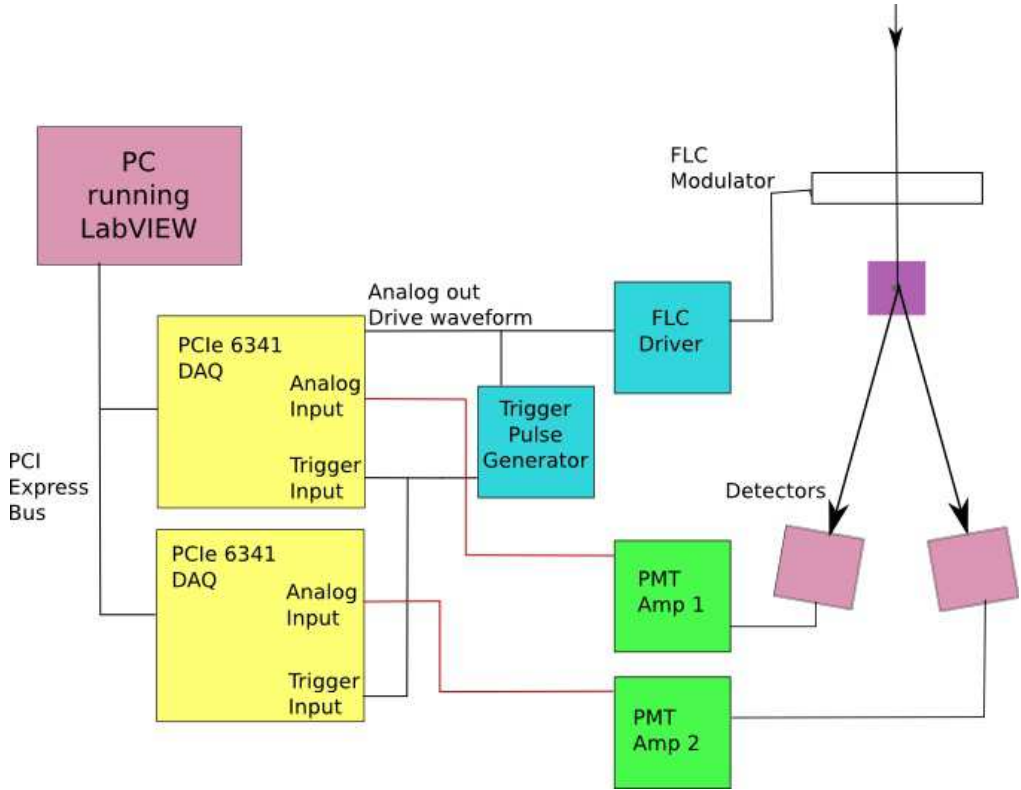}
\caption{Schematic diagram of HIPPI data acquistion system}
\label{fig3}
\end{center}
\end{figure*}

\subsection{Instrument Control and Data Acquisition}

HIPPI is controlled by software running on a rack mount computer (Intel quad core i7, 8 GB RAM, 2 1 TB
disks). The computer runs the Windows 7 Professional operating system and the software has been developed using the
National Instruments LabVIEW graphical programming environment.

The interface to drive the FLC and read data from the detectors makes use of two National Instruments
data acquisition modules (PCIe 6341) each of which provides 16-bit analog input and output
channels. The drive waveform for the modulator is generated in software and output from one of the
modules. A trigger digital input signal is generated from the rising edge of the square wave and fed
to both modules. The detector signals are read by analog input channels on each module. A schematic
diagram of the system is shown in figure \ref{fig3}.

The HIPPI software system also provides control of the FLC temperature, the filter wheel selection,
the rotation of the Wollaston prism and detector section of the optics, and the gain and HT voltage
settings for the detectors.

In operation the analog input channels are sampled at 10 microsecond intervals and read for an integration
time of typically 1 second resulting in 100000 data points for each channel. The timing is
controlled to start on the trigger input so that the sampling always has the same phase relationship
with the modulator waveform. The data are then folded in software over the modulation cycle. For our
standard 500 Hz modulation this results in an array of 200 points from each channel. 

\begin{figure}
\begin{center}
\includegraphics[scale=0.4, angle=0]{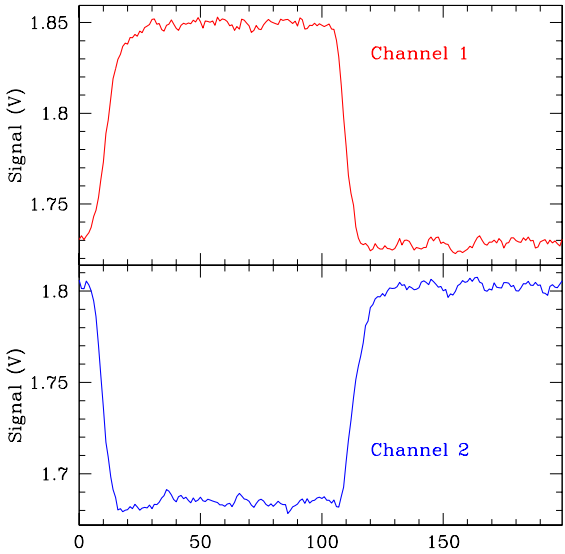}
\caption{Example of the modulated signal seen by HIPPI when observing a polarized star (in this case
HD147084 with about 3.2\% polarization). This is the observed signal integrated over one second and
folded over the 500 Hz modulation cycle giving 200 points over the modulation cycle. The rms noise on each point
is 2.1 mV giving a S/N of 850.}
\label{fig4}
\end{center}
\end{figure}

Figure \ref{fig4} shows what the resulting waveforms look like when a polarized star is being
observed. The modulation is of opposite sign in the two channels as they correspond to the two
orthogonal polarization states from the Wollaston prism. Zero on the diagram corresponds to the rising
edge of the square wave drive signal to the FLC. The delay of about 100 $\mu$s in the observed signal is
a combination of the finite switching time of the FLC, and the time constant (22 $\mu$s) of the detector
amplifiers.

\begin{figure}
\begin{center}
\includegraphics[scale=0.4, angle=0]{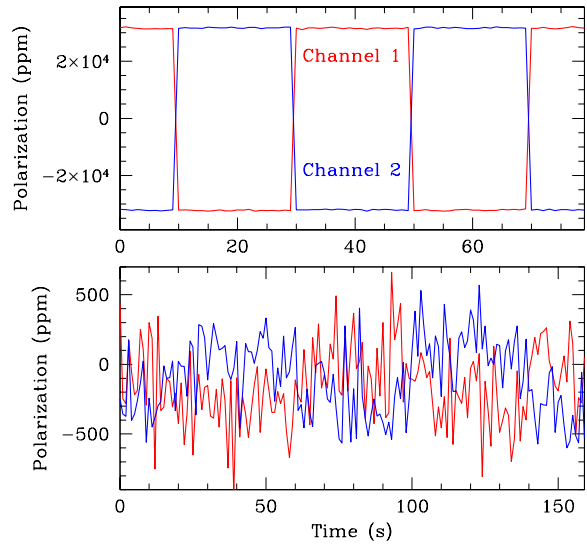}
\caption{Second stage chopping procedure showing the sign of the measured modulation reversing for
each 90 degree rotation of the Wollaston prism and detectors. The top panel is for the highly
polarized star HD 147084 (about 3.2\% polarization in this case), and the lower panel is for a lower
polarizaton object (about 150 ppm, or 0.015 \%). }
\label{fig5}
\end{center}
\end{figure}

These modulation waveforms are displayed to the user and output to the data files every integration. The amplitude of
the modulation is a measure of the polarization of the source. A quick-look data reduction capability
is built into the observing software and derives the fractional polarization (p) as follows.

\begin{equation}
\label{eqna}
p = \frac{(X-Y)}{(X+Y)}
\end{equation}

where X is the signal from the first flat part of the waveform (points from about 20 -- 100 in figure
\ref{fig4}), and Y is the signal from the second flat section (about 120 -- 200), with points around
the two transitions being ignored.  This simple quick-look reduction allows immediate evaluation of incoming
data, but excludes some corrections such as subtracting the small bias and dark signals. 
The offline data reduction described in section \ref{sec_dr} uses the full waveform and inlcudes all 
these corrections.

\subsection{Observing Procedure}
\label{sec_obs}

The normal observing procedure with HIPPI is to make a sequence with the Wollaston
and detectors rotated to two positions 90 degrees apart (referred to as A and B), in the cycle A, B, B, A with this sequence
then being repeated as many times as required. This rotation causes the sign of the modulation in
each channel to reverse. It results in the p value obtained for each integration (with Equation
\ref{eqna} above) following the sequence as shown in figure \ref{fig5}.

Note that while the polarization changes are almost symmetric about zero for the highly polarized star
in the top panel, a significant offset from zero is apparent in the bottom panel. 

An observation of this type measures only one Stokes parameter of linear polarization. To get the
orthogonal Stokes parameter the entire instrument is rotated through 45 degrees and the sequence
repeated. The rotation is performed using the AAT's Cassegrain instrument rotator. In practice we
also repeat the observations at telescope position angles of 90 and 135 degrees to allow removal of
instrumental polarization.

\section{Data Reduction}
\label{sec_dr}

The HIPPI data processing and calibration procedure has at its heart a Mueller matrix model of the
instrument. The parameters of the Mueller matrix are varied according to the state of the system. The
system parameters vary over a single integration as a result of the varying voltage applied to the
FLC. 

The Mueller matrix $\mathbf{M}$ of an optical system (such as HIPPI) relates the output Stokes vector
$\mathbf{s_{out}}$ to  the input Stokes vector $\mathbf{s_{in}}$ through: 

\begin{equation}
\label{eqn_mull}
\mathbf{s_{out}} = \mathbf{M} \mathbf{s_{in}}
\end{equation}

The Mueller matrix for HIPPI is not constant since it changes through the cycle of the modulator. We can 
describe a full integration using a system matrix $\mathbf{W}$. The system matrix is an N by 4 matrix, where each
row is a  state of the system, corresponding to a single data point in the modulation curve. Multiplying the input
Stokes vector by the system matrix gives the vector $\mathbf{x}$ of $N$ observed intensities seen at the detector during the 
modulation cycle (e.g. the $N$ points plotted in figure \ref{fig4}, where $N = 200$).

\begin{equation}
\label{eqn_sys}
\mathbf{x} = \mathbf{W} \mathbf{s_{in}}
\end{equation}

It can be seen that the N rows of the system matrix $\mathbf{W}$ are simply the top rows of the Mueller matrices
$\mathbf{M}$ for HIPPI for each of the N states of the instrument through a modulation cycle. Only the top
row is needed because this is the row of the Mueller matrix that determines the intensity of the output light, and
only intensity is directly measured at the detector. 

The optical components HIPPI is made up of are the FLC and the Wollaston prism. the FLC is a retarder, 
which has the Mueller matrix:
\begin{equation}
\mathbf{M_{ret}} = \begin{bmatrix} 1 & 0 & 0 & 0 \\ 
0 & C^2+S^2 \cos{\delta} & SC(1-\cos{\delta}) & -S \sin{\delta} \\
0 & SC(1-\cos{\delta}) & S^2 + C^2\cos{\delta} & C \sin{\delta} \\
0 & S \sin{\delta} & -C\sin{\delta} & \cos{\delta} 
\end{bmatrix}
\end{equation}
where $C = \cos{2 \phi}$, $S = \sin{2 \phi}$, $\delta$ is the retardance, and $\phi$ the angle 
to the fast axis of the retarder. For a half wave plate the retardance is
$\pi$ radians. FLCs are sometimes modelled as having a linear depolarization component \citep{gendre10}, 
with a Mueller matrix of:
\begin{equation}
\mathbf{M_{Depol}} = \begin{bmatrix} 1 & 0 & 0 & 0 \\
0 & 1-d & 0 & 0 \\
0 &  0 & 1-d & 0 \\
0 & 0 & 0 & 1-d 
\end{bmatrix}
\end{equation}
where $d$ is the degree of depolarization. 

The Wollaston prism behaves as two perpendicular polarizers. A polarizer has a Mueller matrix of:
\begin{equation}
\mathbf{M_{Pol}} = 1/2 \begin{bmatrix} 1 & eC & eS & 0 \\
eC & eC^2 & eSC & 0 \\
eS & eCS & eS^2 & 0 \\
0 & 0 & 0 & 0 
\end{bmatrix}
\end{equation}
where $C = \cos{2 \vartheta}$, $S = \sin{2 \vartheta}$, $e$ is the efficiency of the polarizer, and
$\vartheta$ is the angle of the polarizer axis to that defined for the incoming beam. Wollaston prisms
are very efficient, and we assume $e = 1$.

The combined Mueller matrix for HIPPI can now be written in terms of these matrices as:
\begin{equation}
\mathbf{M} = \mathbf{M_{Pol} M_{Ret} M_{Depol}}
\end{equation}
 and can be used to derive the system matrix for HIPPI as already described. The N rows of the system matrix
are derived from the Mueller matrices for the state of the system at each modulation point
where each row differs due to different values of $\phi$ (in $\mathbf{M_{Ret}}$) and $d$ (in
$\mathbf{M_{Depol}}$).

The other main parameter in the Mueller matrix is the retardance $\delta$. This does not vary around the
modulation cycle, but it is a strong function of wavelength. The retardance is close to half-wave only at a single
wavelength. The effect of a retardance that is not exactly half-wave is to reduce the amplitude of the modulation
curve by a factor of $(1 - \cos{\delta})/2$. Incorporating the full wavelength dependence of retardance in our
reduction model would be impracticable as a single observation can cover a wide range of wavelengths. Instead we
perform the reduction assuming a half-wave retardance, and then account for the wavelength dependence effects by
scaling the polarization with an efficiency correction that can be determined empirically from standard star
observations (section \ref{sec_polstd})  or can be derived from a model (section \ref{sec_bandpass}).

Once the system matrix is knwn equation \ref{eqn_sys} can be inverted \citep{sabatke00} to give:

\begin{equation}
\label{eqn_red}
\mathbf{s_{in}} = \mathbf{W}^+ \mathbf{x}
\end{equation}

where $\mathbf{W}^+$ is the pseudo-inverse of $\mathbf{W}$, which we calculate numerically using the
method of \citet{rust66}. This gives the source Stokes parameters $\mathbf{s_{in}}$ in terms of the observed 
modulation data $\mathbf{x}$.

In practice, because HIPPI isn't designed to measure circular polarization, we use only the first 
three columns of $\mathbf{W}^+$, to obtain a least squares estimate of the components of $\mathbf{s_{in}}$, (
$I$, $Q$ and $U$), in the manner of
\citet*{gendre10} and then divide $Q$ and $U$ by $I$ for the normalised Stokes parameters. 

\subsection{Calibration}

To apply equation \ref{eqn_red} we need to know how the waveplate angle $\phi$ and the depolarization
factor $d$ vary thrugh the modulation cycle of the FLC as these are the only varying components that
enter into $\mathbf{W}$. This is done using a laboratory calibration procedure where we input known
polarization states into the instrument using a lamp and a rotatable polarizer. Measurements are made
of the full modulation curve with the polarizer stepped through a full rotation in 10 or 20 degree
steps. By fitting a Malus law \citep{hecht01} --- modified to include an intensity offset and a phase
shift --- to the intensity as a function of polarizer
angle we determine $\phi$ and $d$ for each point in the modulation cycle. 

To reduce the effect of noise in the calibration measurements the array of $\phi$ values is smoothed
by replacing the last 50 modulation points in the plateau regions (i.e. the last 50
before the FLC is switched) by their average value. Only the last half of the plateau 
region is used because we have found that the Micron
Technology FLC does not have a stable $\phi$ until this point -- it is still approaching
maximum/minimum. 

The values of $\phi$ and $d$ are determined separately for each of the two PMT channels. This is
to allow for timing differences between the two channels. The depolarization ($d$) is included
primarily to account for the switching part of the modulation cycle, where the switching can occur
faster than the detector response resulting in a reduction in apparent polarization. The results 
are normalized such that it does not result in any scaling of the measured polarization, and therefore 
does not attempt to duplicate the wavelength dependent efficiency correction.

\subsection{Measurements}
\label{sec_meas}

\begin{figure}
\begin{center}
\includegraphics[scale=0.48, angle=0]{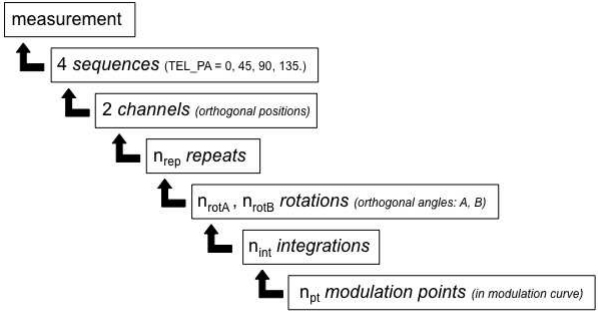}
\caption{The structure of a measurement}
\label{fig6}
\end{center}
\end{figure}

For a given filter and target, a measurement involves taking data at 4 position angles of the
Cassegrain rotator corresponding to 0, 45, 90 and 135 degrees. In general, a sky measurement is 
also made at each
telescope position angle. For very bright targets or high polarization standards or 
bright targets in
moonless conditions a dark measurement may be substituted for the sky measurements. Each sequence is
made up of a number of identical repeats each of which consist of a subsequence of a number of
rotations; for each rotator position there are a number of integrations, each of which consists of
the individual modulation points that make up the modulation curve.  The rotation subsequence is
typically an ABBA sequence with A and B corresponding to orthogonal rotator positions as described in
section \ref{sec_obs}. Figure
\ref{fig6}
summarises the components of a measurement.

As an example, a typical bright star measurement with HIPPI consists of sequences at 4 position
angles each consisting  of 2 repeats ($n_{rep} = 2$), in an 
ABBA rotation subsequence ($n_{rotA} = n_{rotB} = 2$), each
consisting of 10 1 second integrations ($n_{int} = 10$) of a modulation curve 
consisting of 200 modulation points ($n_{pt} = 200$). 
Measurements requiring higher precision have higher numbers of integrations and repeats.

\subsection{Statistical treatment of values and errors}
\label{sec_stats}

The Mueller matrix model of the modulation curve is used to calculate a Q/I and U/I for each
integration. However, the co-ordinate frame of the instrument is chosen to match the calibrated
rotator positions (see section \ref{sec_instpol}, which means that the majority of the modulation points in the curve measure predominantly
either Q or U; the other Stokes parameter determination is discarded because it contributes greater
noise. The ``off-axis'' Stokes parameter is measured by significantly fewer modulation points, and is
contaminated by the intrinsic FLC polarization. This gives a single Stokes parameter determination for each
integration. 

The average of all the integrations for a given rotator position is then calculated, and the error calculated
from the standard deviation of the $n_{int}$ points. The average polarization over each rotation, 
repeat and channel in the sequence
is calculated, and the error on this calculated by averaging the statistical errors for each 
individual rotator position and divding by the square root of the number of these that have been
combined. This gives a normalised Stokes parameter measurement for each telescope position angle
($S_0, S_{45}, S_{90}, S_{135}$), each with an associated error ($E_0, E_{45}, E_{90}, E_{135}$).

The value of $Q/I$, in the chosen co-ordinate system of the instrument, is obtained from the average of
the 0 and 90 degree sequences and the value of $U/I$ from the average of the 45 and 135 degree
sequences. 

\begin{equation}
Q/I = \frac{S_0 + S_{90}}{2}
\end{equation}
\begin{equation}
U/I = \frac{S_{45} + S_{135}}{2}
\end{equation}

With errors calculated as:

\begin{equation}
E_{Q/I} = \frac{E_0 + E_{90}}{2\sqrt{2}}
\end{equation}
\begin{equation}
E_{U/I}  = \frac{E_{45} + E_{135}}{2\sqrt{2}}
\end{equation}

\subsection{Sky and dark correction}

Sky measurements are made to take account, predominantly, of polarization caused by reflected light
from the Moon. The sky sequences are always made with the same HT voltage, gain, telescope position
angle and filter as the science sequences. Because the sky signal level is much lower than that from
the star, much shorter total integration times can be used for the sky measurements without
intoducing significant additional noise. The average modulation curves for the sky data are determined
for each PMT channel and rotator position. These are then subtracted from the each of the
corrsesponding modulation curves in the science sequence prior to the analysis described in sections
\ref{sec_meas} and \ref{sec_stats}.

In cases where the sky contribution is negligible, for example when observing a bright star in
moonless conditions, a dark observation can be used as an alternative to a sky. This is made by using
a blank in the filter wheel to exclude any light input to the instrument.

\begin{table*}
\caption{Effective wavelength and modulation efficiency for different spectral types according to bandpass model}
\label{tab_sptype}
\begin{tabular}{lllllll}
\hline Spectral Type & \multicolumn{3}{c}{Effective Wavelength (nm)} & \multicolumn{3}{c}{Modulation Efficiency (\%)} \\
      &    500SP & g$'$ & r$'$ & 500SP & g$'$ & r$'$ \\   \hline
B0 V  &  433.2  &  472.6  &  600.8  &  74.3  &  90.8  &  84.4 \\
A0 V  &  443.7  &  475.6  &  601.9  &  81.6  &  91.5  &  84.2 \\
F0 V  &  447.1  &  479.5  &  603.7  &  83.0  &  92.3  &  83.8 \\
G0 V  &  451.2  &  483.4  &  605.5  &  84.6  &  93.0  &  83.5 \\
K0 V  &  456.4  &  486.4  &  606.6  &  87.1  &  93.7  &  83.3 \\
M0 V  &  459.2  &  490.1  &  610.5  &  88.8  &  93.8  &  82.6 \\
M5 V  &  458.1  &  490.0  &  611.4  &  88.3  &  93.7  &  82.4 \\
\hline
\end{tabular}
\end{table*}

\subsection{Coordinate transformation and efficiency calibration}

The co-ordinate transformation from the frame of the instrument to that of the sky is carried out by
reference to known polarization standards. Polarization angle is calculated by:

\begin{equation}
\theta = 1/2 \arctan{\frac{U/I}{Q/I}}
\end{equation}

The difference between the calculated angle with that known for the standards becomes the angular
correction, and all other measurements are rotated by this value.

The linear polarization, P, is given by:

\begin{equation}
P = \sqrt{(Q/I)^2 + (U/I)^2}
\end{equation}

This polarization then needs to be scaled by multiplying by an efficiency correction to account for the wavelength
dependence of the retardance of the modulator. This efficiency correction can either be determined empirically
using polarized standard stars as described in section \ref{sec_polstd}, or determined using the bandpass model
described in section \ref{sec_bandpass}.

\subsection{Filters and bandpass model}
\label{sec_bandpass}

The filters used in HIPPI are relatively broad (150 nm for the g$'$ and r$'$ filters) and hence the precise effective
wavelength will vary with star colour and other factors, and this will, in turn, affect the modulation efficiency which is
also a function of wavelength since the modulator is not an achromatic device. To account for these effects we use a
bandpass model similar to that described by \citet{hough06}. 

As a starting point for such a model we typically use one of the \citet{castelli04} stellar atmosphere models. The
spectral energy distribution (SED) given by the model can then be modified for interstellar extinction using the
empirical model of \citet{cardelli89}. We then correct the SED for its passage through the Earth's atmosphere by
applying an atmospheric transmission correction calculated using the VSTAR modelling code \citep{bailey12} and 
including molecular absorption and Rayleigh scattering.

Finally this can be combined with a model of the instrument response which includes the transmission functions of the
filters, and the cathode radiant sensitivity (in mA/W) of the PMT as taken from the Hamamatsu data sheet. The final
result, which we call $S(\lambda)$ is the relative contribution to the output detector signal as a function of wavelength.

The effective wavelength of the observation can be determined from

\begin{equation}
\lambda_{eff} = \frac{\int \lambda S(\lambda) d\lambda}{\int S(\lambda) d\lambda}
\end{equation}

where the integral is taken over all wavelengths for which $S(\lambda)$ is non-zero.

The polarization modulation efficiency correction will be given by

\begin{equation}
P_c = \frac{\int e(\lambda) S(\lambda) d\lambda}{\int S(\lambda) d\lambda}
\end{equation}

where $e(\lambda)$ is the wavelength dependence of the modulation efficiency. For an FLC the efficiency as a
function of wavelength is primarily determined by the variation of the retardance with wavelength. According to 
\citet*{gisler03} the optical path difference ($\Delta$) of an FLC can be represented by:

\begin{equation}
\Delta = \frac{\lambda_0}{2} + Cd \left(\frac{1}{\lambda^2} - \frac{1}{\lambda_0^2}\right)
\end{equation}

where $\lambda_0$ is the half-wave wavelength, $C$ is a parameter describing the dispersion in birefringence of
the FLC material, and $d$ is the thickness of the FLC layer. In practice $Cd$ can be treated as a single
parameter. We have determined these parameters for our two modulators by fitting this formula to a set of
laboratory measurements of a polarized source made with HIPPI using a set of narrow band filters.
The results are given in table \ref{tab_mod_props}.

\begin{table}
\caption{Properties of FLC Modulators}
\label{tab_mod_props}
\begin{tabular}{lll}
\hline
Modulator & Micron & BNS \\  \hline
$\lambda_0$ & 505 $\pm$ 5 nm & 498 $\pm$ 5 nm\\
$Cd$ & $1.75 \pm 0.05 \times 10^7$ nm$^3$ & $1.70 \pm 0.05 \times 10^7$ nm$^3$ \\
\hline
\end{tabular}
\end{table}

The properties of the two modulators are very similar, and the $Cd$ value we obtain is almost the same as that found by
\citet{gisler03} for a similar device.

The modulation efficiency is then given by:

\begin{equation}
e(\lambda) = \frac{e_{max}}{2}\left(1 - \cos{2 \pi \frac{\Delta}{\lambda}}\right)
\end{equation}

where $e_{max}$ is the peak efficiency measured at wavelength $\lambda_0$. Ideally $e_{max}$ should equal 1, but
we find a value of 0.98 is the maximum achievable using laboratory measurement with HIPPI. 

\begin{figure*}
\begin{center}
\includegraphics[scale=0.4, angle=0]{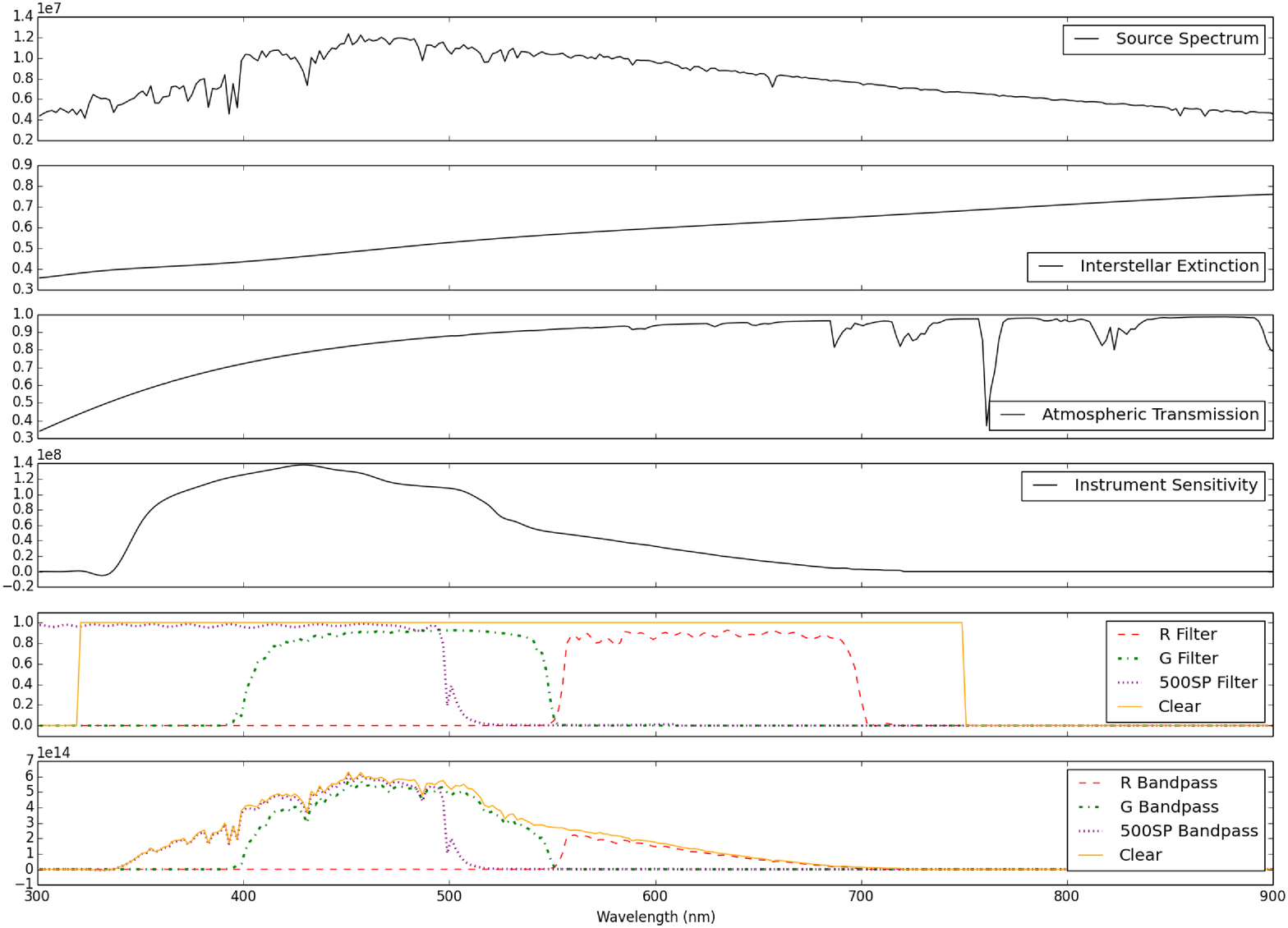}
\caption{Illustration of the various components that combine to give the effective filter bandpass
($S(\lambda)$) shown in the lower panel of the plot}
\label{fig_bandpass}
\end{center}
\end{figure*}

Table \ref{tab_sptype} shows how the effective wavelength and polarization efficiency vary with spectral
type. The bandpass model is illustated in figure \ref{fig_bandpass}.

\section{Performance and Results}

The performance of HIPPI has been evaluated based on our observing runs on the AAT over May 8 -- 12
2014 and Aug 28 -- Sep 2 2014.

\subsection{Instrumental polarization}
\label{sec_instpol}

We have found that the FLC modulators used in HIPPI introduce a large instrumental polarization
effect which appears to be an intrinsic property of the modulator. Such polarization is thought
to arise from multiple internal reflections in the birefringent material between the plates resulting
in wavelength dependent fringe patterns in transmittance that are different for light polarized
parallel or perpendicular to the fast axis \citep{gisler03,ovelar12}.
This effect was found to be
present with both our modulators but is a factor of about 3 higher for the Micron Technology modulator
(used for the May 8 -- 12 observing run) than for the BNS modulator (used for the Aug 28 -- Sep 2
run). Even in the BNS case the effect is at the $\sim$1000 ppm level and highly variable with
wavelength. Since it is due to the modulator that does not rotate, it cannot be removed by our second-stage
chopping procedure described earlier.

Fortunately because the instrument only measures one Stokes parameter at a time it is possible to
arrange things such that the instrumental polarization is orthogonal to the Stokes parameter being
measured. This is done by choosing optimal values for the pair of 90 degree separated angles that are
used for rotating the Wollaston prism and detectors. When this is done the instrumental polarization
is greatly reduced but it cannot be removed completely. Due to effects such as its wavelength
dependence in angle, residual instrumental effects at around the 50 ppm level remain. 

The residual instrumental polarization is removed by the procedure of observing at four position
angles of the instrument rotator (0, 45, 90, 135) as described in sections \ref{sec_obs} and
\ref{sec_meas}. The 90 degree rotation between the 0, 90 and 45, 135 pairs reverses the sign of the
instrumental polarization relative to that from the star, enabling it to removed.

\subsection{Telescope polarization}

\begin{table}
\caption{Low polarization star measurements to determine telescope polarization}
\begin{tabular}{lllll}
\hline Star & Date & Filter & P (ppm) & $\theta$ (deg) \\  
\hline BS 5854 & Aug 28 & g' & 35 $\pm$ 5 & 113 $\pm$ 6 \\
        & Aug 29 & g$'$ & 46 $\pm$ 4 & 115 $\pm$ 17 \\
	& Aug 30 & g$'$ & 50 $\pm$ 4 & 114 $\pm$ 4 \\
	& Sep 2 &  g$'$ & 45 $\pm$ 5 & 110 $\pm$ 6 \\
	& Average & g$'$ & 44 $\pm$ 2 & 114 $\pm$ 3 \\
	\\
Beta Hyi & Aug 28 & g$'$ & 52 $\pm$ 3 & 113 $\pm$ 4 \\
         & Aug 29 & g$'$ & 62 $\pm$ 3 & 106 $\pm$ 3 \\
	 & Aug 30 & g$'$ & 58 $\pm$ 3 & 109 $\pm$ 4 \\
	 & Aug 31 & g$'$ & 52 $\pm$ 3 & 106 $\pm$ 4 \\
	 & Average & g$'$ & 56 $\pm$ 2 & 109 $\pm$ 2 \\
	 \\
Sirius   & Aug 31 & g$'$ & 55 $\pm$ 1 & 112 $\pm$ 1 \\
         & Sep 2 & g$'$ &  48 $\pm$ 4 & 108 $\pm$ 5 \\
	 & Average & g$'$ & 51 $\pm$ 2 & 110 $\pm$ 3 \\
	 \\
Adopted TP & &      g$'$  & 48 $\pm$ 5 & 111 $\pm$ 2 \\
\hline
\end{tabular}
\label{tab_tp}
\end{table}

The telescope optics can be expected to introduce a small telescope polarization (TP) that must be corrected for. In
the case of PlanetPol on the William Herschel Telescope TP was found to be around 15 ppm
\citep{hough06}. A much
larger TP of $\sim$ 250 ppm was found using POLISH on the 5-m Hale Telescope \citep{wiktorowicz08}.

\begin{table*}
\begin{center}
\caption{Predicted polarization in the HIPPI filters for polarized standard stars}
\begin{tabular}{llllllllllll}
\hline Star & Spectral Type & $E_{B-V}$ & $R_V$ & $P_{max}$ & $\lambda_{max}$ & K & $\theta$ & \multicolumn{3}{c}{Expected
$P$ (\%)} & Refs \\
 & & & & & & & & g$'$ & r$'$ & 500SP & \\  \hline
HD 23512 & A0 V & 0.34 & 3.2 & 2.29 & 0.60 & 1.06 & 29.9 & 2.15 & & & 1, 2 \\
HD 147084 & A4 II/III & 0.72 & 3.9 & 4.34 & 0.67 & 1.15 & 32.0 & 3.77 & 4.26 & 3.60 & 3, 4 \\
HD 187929 & F6-G0 I & 0.18 & 3.1 & 1.76 & 0.56 & 1.15 & 93.8 & 1.70 & & 1.66 &  5 \\
\hline
\end{tabular}

References:   1.  \citet{guthrie87}, 2.    \citet{hsu82},  3.   \citet{wilking80},   
4.  \citet{martin99},   5. \citet*{serkowski75}  
\label{tab_stand}
\end{center}
\end{table*}

\begin{table*}
\caption{Observations of polarized standard stars}
\label{tab_polstd}
\begin{tabular}{lllllllll}
\hline Star & Date & Filter & \multicolumn{2}{c}{Measured} & \multicolumn{2}{c}{Expected} & Efficiency & Efficiency\\
            &      &        &   $P$ (ppm)  &  $\theta$ (deg) &  $P$ (ppm)  &  $\theta$ (deg) & (Observed) & (Modelled)\\
\hline HD 23512 & Aug 28 & g$'$ & 18852 $\pm$ 37 & 30.5 $\pm$ 0.1  & 21469 & 29.9 &  87.8 & 90.0  \\
\\
HD 147084 & Aug 29 & g$'$  &  33919 $\pm$ 11  &  32.0 $\pm$ 0.1  & 37664  & 32.0 & 90.1 & 91.0 \\
         & Aug 30 & g$'$  &  33972 $\pm$ 12  &  32.0 $\pm$ 0.1  &  37665 & 32.0 & 90.2  & 91.0 \\
	 & Aug 30 & r$'$  &  34910 $\pm$ 17  &  32.1 $\pm$ 0.1  &  42619 & 32.0 & 81.9 & 84.2 \\
	 & Aug 30 & 500SP & 29349 $\pm$ 17  &  32.0 $ \pm$ 0.1 & 35958 & 32.0 & 81.6 & 80.7 \\
	 & Aug 31 & 500SP & 29337 $\pm$ 12  &  31.9 $ \pm$ 0.1 & 35956 & 32.0 & 81.6 & 80.7 \\
	 \\
HD 187929 & Aug 28 & g$'$ &   15450 $\pm$ 9 &  93.5 $\pm$ 0.1 & 16950  &  93.8  & 91.1 & 91.2 \\
          & Aug 29 & g$'$ &   15493 $\pm$ 9 &  93.6 $\pm$ 0.1 & 16960  &  93.8  & 91.3 & 91.4 \\
	  & Sep 2  & g$'$ &   15560 $\pm$ 13 & 93.6 $\pm$ 0.1 & 16944  &  93.8  & 91.8 & 91.1 \\
	  & Sep 2  & 500SP & 14213 $\pm$ 16 & 93.8 $\pm$ 0.1 & 16570   &  93.8 & 85.7 & 84.7 \\
\hline
\end{tabular}
\end{table*}

As the AAT is an equatorially mounted telescope it is not possible to separate the telescope and star
polarization in the way described by \citet{hough06} making use of the field rotation in an
altazimuth mounted telescope.

Instead, we have adopted as our preliminary estimate of the telescope polarization the average of our
observations of three stars which we have reason to believe should have very low polarization. One of
these is BS 5854 which is one of the low polarization standard stars observed with PlanetPol. It is
at a distance of 22.5 pc and had measured polarizations with PlanetPol of 5 ppm or lower
\citep{hough06,bailey08}. The other two are nearby bright stars Beta Hydri (HIP 2021, BS 98) which is
a 2.8 magnitude star at a distance of 7.5 pc, and Sirius ($\alpha$ CMa) at a distance of 2.6 pc.
Based on the polarization with distance results of \citet{bailey10} these should be expected to have
very low polarizations.

The results are given in table \ref{tab_tp} for the telescope polarization measurements in the SDSS g$'$
filter. From smaller numbers of observations we have determined TP = 53 $\pm$ 2 ppm, $\theta$ = 110 
$\pm$ 2 degrees in the 500SP filter, TP = 45 $\pm$ 3, ppm $\theta$ = 125 $\pm$ 5 degrees in the SDSS r$'$ filter, and TP =
49 $\pm$ 2 ppm, $\theta$ = 117 $\pm$ 2 degrees with no filter (the full wavelength range of the detector). These
results suggest an increasing TP and a rotation in angle from red to blue.

\subsection{Polarized Standard Stars}
\label{sec_polstd}

Observations were made of a number of polarized standard stars. The stars observed are listed in
table \ref{tab_stand}, together with the parameters needed for the bandpass model.

The wavelength dependence of interstellar polarization can be represented by the empirical model
\citep{serkowski75,wilking80}

\begin{equation}
P(\lambda) = P_{max} \exp \left( -K \ln^2 \frac{\lambda}{\lambda_{max}} \right)
\end{equation}
 
where the values of $P_{max}$, $\lambda_{max}$ and $K$ are empirical parameters that are fitted to
observed wavelength dependent polarization measurements and are given in table \ref{tab_stand} for our
standard stars. 

The expected polarization in any of the HIPPI filters can then be predicted by averaging this function over the
bandpass model as described in section \ref{sec_bandpass} as follows:

\begin{equation}
P_p   = \frac{\int P(\lambda) e(\lambda) S (\lambda) d \lambda}{\int e(\lambda) S(\lambda) d\lambda}
\end{equation}

Observations of polarized standard stars are listed in table \ref{tab_polstd}.

It can be seen from table \ref{tab_polstd} that the repeatability of these measurements from night to
night is excellent with polarization agreeing to 0.01 \% (100 ppm) or better and the position angles
agreeing to typically 0.1 deg. The position angles also agree quite well with the expected values. While
the position angle zero point was calibrated using these stars, the agreement between the three stars is
an indicator of the accuracy of these measurements. The absolute calibration of position angle is
somewhat poorer than the internal errors of $\sim$0.1 degree, because it is limited by the available data
on standard stars.

Comparing with the predicted values of polarization from table \ref{tab_stand} we find that the observed
modulation efficiency averages $\sim$90 \% in the g' band with somewhat lower values in the 500SP and r'
filters. These can be compared with the predicted efficiencies from the model described in section
\ref{sec_bandpass} which are given in the final column of the table. It can be seen that the observed efficiencies are in good agreement with the efficiencies
predicted by our model. In most cases the measured and observed values agree within 1\%, with the largest
deviation being 2.3 \%. Given that the quoted accuracies of the standard star values are typically around 1\%
of the measured value this is excellent agreement.

\subsection{Polarization Sensitivity}

The polarization sensitivity of a polarimeter is its ability to measure small polarization levels, and is equivalent
to the precision of measurement on low polarization objects. It can be estimated by looking at the night to night
scatter of repeat observations of low polarization stars. In table \ref{tab_sens} we show such results for four sets
of observations on three stars that have been measured on three or more nights. The data presented here are the instrumental 
Stokes parameters as measured by HIPPI
before corrections for telescope polarization, and position angle zero point.

\begin{table*}
\caption{Repeat observations of low polarization objects to determine sensitivity}
\label{tab_sens}
\begin{tabular}{lllllllll}
Star & \multicolumn{2}{c}{Canopus (g$'$)} & \multicolumn{2}{c}{Beta Hyi (g$'$)} & \multicolumn{2}{c}{BS 5854 (g$'$)} &
\multicolumn{2}{c}{Beta Hyi (500SP)} \\
 & Q/I (ppm) & U/I (ppm) & Q/I (ppm) & U/I (ppm) & Q/I (ppm) & U/I (ppm) & Q/I (ppm) & U/I (ppm) \\  \hline
 & 130.7 $\pm$ 0.8 & $-$62.7 $\pm$ 0.8 & 42.1 $\pm$ 3.2 & $-$22.7 $\pm$ 3.1 & 28.5 $\pm$ 4.8 & $-$14.5 $\pm$ 2.4 & 38.5 $\pm$ 3.4 & $-$27.2 $\pm$ 3.4 \\
 & 134.4 $\pm$ 1.2 & $-$64.9 $\pm$ 1.8 & 42.2 $\pm$ 3.2 & $-$37.8 $\pm$ 3.2 & 37.7 $\pm$ 3.6 & $-$17.5 $\pm$ 13.8 & 41.0 $\pm$ 3.3 & $-$27.6 $\pm$ 3.3 \\
 & 129.7 $\pm$ 1.3 & $-$58.4 $\pm$ 1.3 & 43.1 $\pm$ 3.3 & $-$31.3 $\pm$ 3.1 & 41.2 $\pm$ 3.6 & $-$19.9 $\pm$ 3.3 & 35.2 $\pm$ 3.5 & $-$19.6 $\pm$ 3.6 \\
 & 128.7 $\pm$ 3.1 & $-$58.7 $\pm$ 2.5 & 35.0 $\pm$ 3.4 & $-$32.4 $\pm$ 3.1 & 33.5 $\pm$ 5.1 & $-$23.3 $\pm$ 3.6 & & \\
STDEV & 2.5  &  3.2  &  3.8 &  6.2 & 5.5 & 3.7 & 2.9 & 4.5 \\
\hline
\end{tabular}
\end{table*}

The STDEV row at the bottom of the table gives the standard deviation of the numbers in the column and measures that
night-to-night scatter in the repeat observations. The average of these values is 4.0 ppm, equivalent to
4.3 ppm when the modulation efficiency correction is included.

While this represents our current best estimate of the sensitivity achieved with HIPPI, it should be
noted that the internal errors of most of these observations are typically 3 -- 5 ppm, so part of the
scatter may well be due to the contribution of random noise. For Beta Hydri (B = 3.4) the expected photon rate (above
the atmosphere) at the AAT in the g$'$ band is $1.06 \times 10^{10}$ photons sec$^{-1}$. Assuming a 10\% total throughput
allowing for atmospheric, telescope, and instrument losses as well as the detector QE (about 25\% averaged over the band),
we detect a total of 1.92 x 10$^{11}$ photons in the 180 second integration used for these observations. The expected photon
shot noise limited sensitivity per Stokes parameter ($1/\sqrt{N_{photons}}$) is then 2.3 ppm. The excess noise factor of
the photomultiplier tubes which is
typically around 1.2, will increase this to $\sim$2.8 ppm. This is close to the measured errors for this object of 3.1 to 3.4 ppm
indicating that even on these bright objects, where the photomultiplier gain is set to its lower values, we are obtaining
close to photon-noise-limited performance.

The true polarization sensitivity achievable
with HIPPI may therefore be somewhat better than these figures indicate. The lower figures for the brightest of these stars (Canopus)
indicate a value nearer 3 ppm.

These figures can be compared with the performance of other ``parts-per-million'' polarimeters. For
PlanetPol, using data from \citet{bailey08} the equivalent sensitivity figure is 2.1 ppm. For POLISH
using data from table 3 of \citet{wiktorowicz08} the figure is 8.5 ppm. The sensitivity of 4.3 ppm
achieved with HIPPI shows that a polarimeter using FLC modulation is quite competitive with the PEM
technology used in these other two instruments.

\section{Conclusions}

We have built and tested a stellar polarimeter based on a ferro-electric liquid crystal (FLC) modulator.
The polarimeter has been used successfully on the Anglo-Australian Telescope. The FLC does not provide as
``polarimetrically clean'' a system as the photo-elastic modulators (PEMs) used in other high precision
polarimeters. Nevertheless, after correction of instrumental effects using multiple stages of modulation,
HIPPI achieves performance comparable with PEM polarimeters such as PlanetPol \citep{hough06}
and POLISH \citep{wiktorowicz08}, while offering advantages of lower cost, compact size and greater
efficiency.

HIPPI provides both high precision --- it can measure the polarization of highly polarized stars to a
precision in fractional polarization of 0.01 \% and position angle to around 0.1 degree or better --- and high sensitivity --- it
can measure very low levels of polarization with a precision on low polarization objects of 4.3 ppm or
better. The wide dynamic range of its detectors allow it to observe objects from the brightest stars in
the sky (e.g. the observations of Sirius and Canopus reported here), while still having the capability to
observe quite faint objects (dark noise from the PMTs should only start to limit performance for stars
fainter than about 16th magnitude). 

We have measured the telescope polarization of the AAT to be 48 $\pm$ 5 ppm in the SDSS g$'$ filter. The
precision of HIPPI is such that the primary limitation on the absolute accuracy of its measurements is
currently set by the lack of suitable standard stars for calibration. There are very few low polarization
stars measured at the parts-per-million level, and this limits our ability to reliably determine and
correct for the telescope polarization. Similarly the precision of our measurements of polarized standard
stars is better than the absolute calibration of their polarizations and position angles currently
available.

\section*{Acknowledgments}

The development of HIPPI was funded by the Australian Research Council through Discovery Projects grant
DP140100121 and by the UNSW Faculty of Science through its Faculty Research Grants program. The authors
thank the Director and staff of the Australian Astronomical Observatory for their advice and  support
with interfacing HIPPI to the AAT and during the two observing runs on the telescope.

\label{lastpage}

\end{document}